\documentclass[%
reprint,
 amsmath,amssymb,
 aps,
]{revtex4-2}

\usepackage{graphicx}
\usepackage{dcolumn}
\usepackage{bm}
\usepackage{graphicx}
\usepackage{makecell}

\begin{document}
\renewcommand{\thetable}{\arabic{table}}
\preprint{APS/123-QED}

\title{Simultaneous inverse-design of material and structure via deep-learning: Demonstration of dipole resonance engineering using core-shell nanoparticles}

\author{Sunae So}
\altaffiliation{Contributed equally to this work}
\affiliation{Department of Mechanical Engineering,
Pohang University of Science and Technology (POSTECH), 77 Chungam-ro, Republic of Korea}

\author{Jungho Mun}
\altaffiliation{Contributed equally to this work}
\affiliation{Department of Chemical Engineering,
Pohang University of Science and Technology (POSTECH), 77 Chungam-ro, Republic of Korea}

\author{Junsuk Rho}
\altaffiliation{Corresponding author : jsrho@postech.ac.kr}
\affiliation{Department of Mechanical Engineering,
Pohang University of Science and Technology (POSTECH), 77 Chungam-ro, Republic of Korea}
\affiliation{Department of Chemical Engineering,
Pohang University of Science and Technology (POSTECH), 77 Chungam-ro, Republic of Korea}

\date{\today}

\begin{abstract}
Recent introduction of data-driven approaches based on deep-learning technology has revolutionized the field of nanophotonics by allowing efficient inverse design methods. In this paper, simultaneous inverse design of materials and structure parameters of core-shell nanoparticle is achieved for the first time using deep-learning of a neural network. A neural network to learn correlation between extinction spectra of electric and magnetic dipoles and core-shell nanoparticle designs, which include material information and shell thicknesses, is developed and trained. We demonstrate deep-learning-assisted inverse design of core-shell nanoparticle for 1) spectral tuning electric dipole resonances, 2) finding spectrally isolated pure magnetic dipole resonances, and 3) finding spectrally overlapped electric dipole and magnetic dipole resonances. Our finding paves the way of the rapid development of nanophotonics by allowing a practical utilization of a deep-learning technology for nanophotonic inverse design.
\end{abstract}

\maketitle

\section{\label{sec:introduction} Introduction}
 Optical interaction between subwavelength particles and light has been an important problem in many fields of physics and materials. However, the subwavelength nanoparticles usually undergo rather uninteresting Rayleigh scattering; on the other hand, subwavelength plasmonic nanoparticles have shown sharp and strong extinction peak due to the localized surface plasmon resonance, which has been applied to refractive index sensing \cite{willets:2007, mock:2003}. The subwavelength plasmonic nanoparticles generally have dominant ED mode with its resonance determined by the plasmonic materials used. Recently, structured subwavelength meta-atoms with higher-order multipole modes have been realized \cite{zhang:2008, kuznetsov:2012}, whose contributions have demonstrated many interesting phenomena including directional scattering \cite{liu:2018}, Fano-like resonance \cite{liu:2017}, and negative index media \cite{paniagua:2011}. Core-shell nanoparticle is a candidate meta-atom to exhibit such exotic phenomena \cite{liu:2012, li:2015, naraghi:2015}, but its higher degree-of-freedom makes designing difficult, and it is usually considered difficult to independently engineer electric dipole (ED) and magnetic dipole (MD) resonances.
    
These design problems due to higher degree-of-freedom can be overcome by utilizing data-driven approaches based on deep-learning (DL).
By training with a vast amount of data, artificial neural networks (NNs) can learn the correlation between various photonic structures and their optical properties, and inversely design complex photonic structures \cite{malkiel2018plasmonic, peurifoy2018nanophotonic, liu2018generative, ma2018deep}. Once trained, a NN can be repeatedly employed to design photonic structures from given arbitrary optical properties, and this design process takes only a fraction of time compared to conventional labor-intensive parameter optimization method. This feature is especially beneficial to computational electromagnetics, which often requires significant computational costs, whereas NN-based design process can be conducted in a single GPU processor.

Nevertheless, to the best of our knowledge, a simultaneous inverse design of structural parameters and material information of photonic structures has not been demonstrated. This difficulty of simultaneous inverse design arises from problem setting, where structural parameters with continuous values should be predicted by \emph{regression problem}, while materials by \emph{classification problem}; therefore, a joint of regression and classification problems should be solved to inversely design material and structural parameters simultaneously. In this study, loss function, which is a function that a NN aims to minimize during a training process, is developed to solve such a multi-task problem \cite{zhu2014novel} consisting of regression and classification problems.
    
In this work, we propose DL-assisted method to inversely design structural parameters and material information of core-shell nanoparticles simultaneously from given ED and MD extinction spectra. Details of the developed NNs and loss function are discussed. To discuss practical usage of DL-assisted inverse design, three cases of dipole resonance engineering are demonstrated: 1) spectral tuning ED resonances, 2) finding spectrally isolated MD resonances, and 3) finding spectrally overlapped ED and MD resonances.

\section{Results and discussion}
\subsection{Deep-learning}
\begin{figure}[h]
    \includegraphics[width=\linewidth]{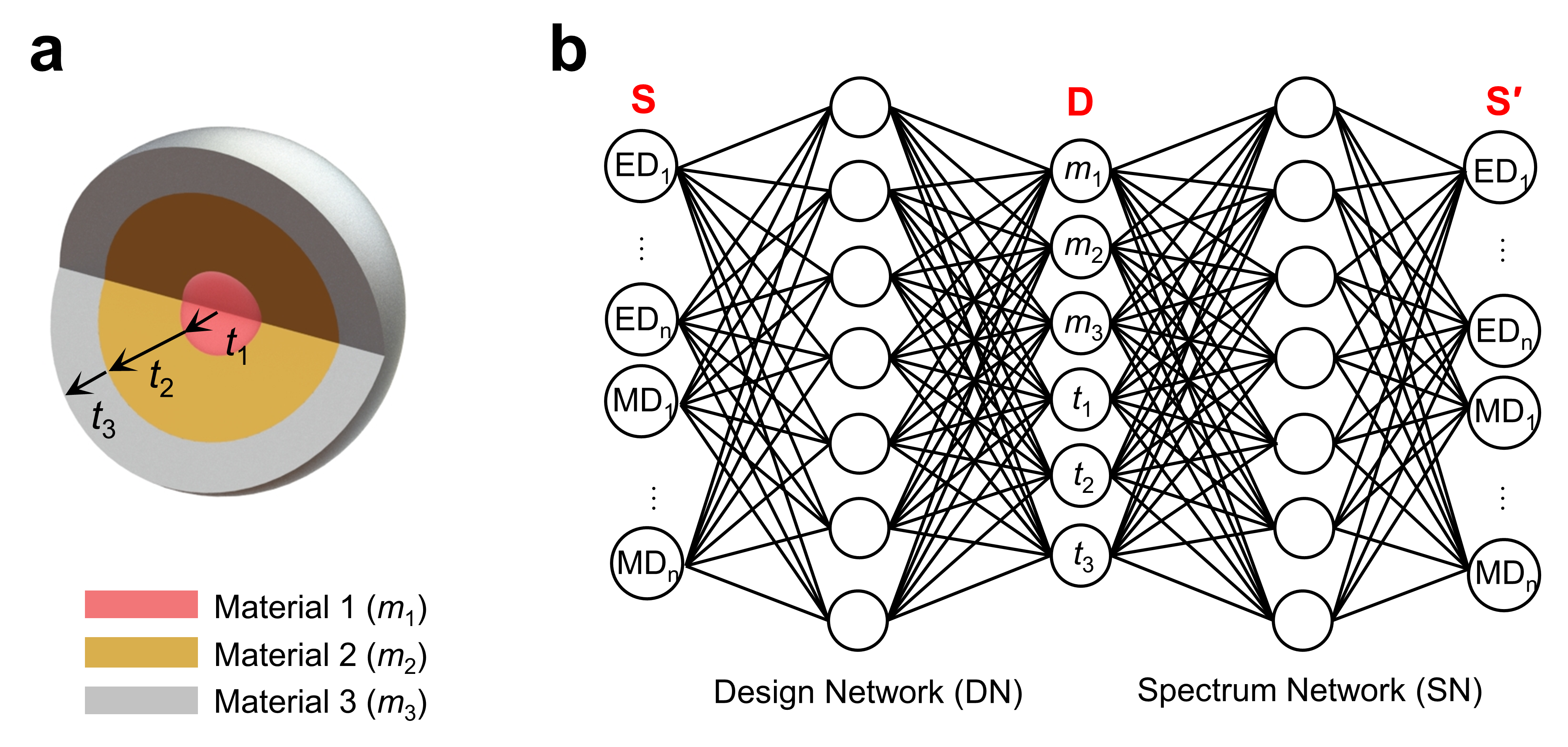}
    \centering
    \caption{Schematics of (a) a core-shell nanoparticle and (b) the utilized DL model consisting of two NNs (DN and SN). DN learns mapping from $\textbf{S}$ (input extinction spectra) to $\textbf{D}$ (design parameters), and SN learns mapping from $\textbf{D}$ to $\textbf{S'}$.}
    \label{fig:deep}
\end{figure}
    
    In this work, spherical three-layered core-shell nanoparticles are used with parameterized material type and thickness for each layer (Fig.~\ref{fig:deep}a). The outer radius of the core-shell is constrained between 10~nm and 150~nm. The extinction for stratified spherical particles (data-set for DL) was obtained using T-matrix formalism at plane-wave incidence. The extinction spectra are normalized by maximum values for simplicity of inverse design (see Appendix F). Each entry of the data-set consists of six design parameters (three materials and three thicknesses) and 100 spectral points of each of ED and MD extinction spectra at wavelength range from 300~nm to 1000~nm. 
    
    Optical properties of meta-atoms significantly depend on their materials. However, refractive index should not be directly used for DL, because arbitrarily returned refractive index cannot generally be realized with real materials. In addition, plasmonic and many dielectric materials are highly dispersive and cannot easily be treated in DL. Therefore, we have listed the seven possible materials (Ag, Au, Al, Cu, TiO$_2$, SiO$_2$, and Si) for design, and indexed them through numbering (Table~\ref{tab:material-index}).
\begin{table}[h]
    \small
    \begin{tabular}{cccccccc}
    \hline
    0 & 1 & 2 & 3 & 4 & 5 & 6 & 7 \\
    \hline
    None & Ag & Au & Al & Cu & TiO$_2$ & SiO$_2$ & Si \\
    \hline
    \end{tabular}
    \centering
    \caption{\textbf{Indexed materials}}
    \label{tab:material-index}
\end{table}
    
    We now express the central idea of this work: our DL network architecture and newly devised loss function, which enable simultaneous inverse design of material and structural parameters. The DL model used for training consists of two networks: a design network (DN) and a spectrum network (SN) (Fig.~\ref{fig:deep}b). DN learns mapping from optical properties (EDs and MDs) to design parameters (materials and thicknesses); and SN, from design parameters to optical properties. The developed overall loss function of DN ($l_\mathrm{overall}$) is defined as weighted average of design ($l_\mathrm{design}$) and spectrum losses ($l_\mathrm{spectrum}$):
    \begin{equation}
        l_{\text{overall}} = w l_{\text{spectrum}} 
        +(1-w) l_{\text{design}},
    \end{equation}
    where $l_{\text{spectrum}}$ is calculated by mean squared error (MSE) of $\textbf{S}$ (target spectrum) and $\textbf{S'}$ (predicted response by DL). The weight $w$ is chosen by parametric study to minimize MSE between $\textbf{S}$ and $\textbf{S''}$ (calculated spectra using the predicted designs $\textbf{D}$ by DN) for the final test data-set (see Appendix A). Simultaneous inverse design of optical materials and structural parameters requires a development of the loss function, because classification and regression problems should be solved at the same time. The devised design loss function is weighted average of material and structural losses:
    \begin{equation}
        l_{\text{design}} = \rho l_{\text{structure}} +(1-\rho) l_{\text{material}},
        \label{eqn:design loss}   
    \end{equation}
    where $\rho$ is a weight of the structural error which is also added as a hyperparameter to be adjusted. $l_\text{structure}$ was evaluated by mean absolute error, and $l_\text{material}$ by binary cross entropy with logits loss. See Appendix A for detailed expression of loss functions and details of parametric study and DL procedures.
    
\begin{figure}[!ht]
    \centering
    \includegraphics[width=\linewidth]{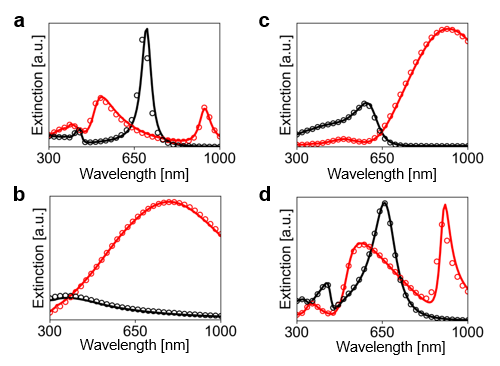}
    \caption{\textbf{Test examples of the inverse design using test data-set.} Spectra of target input (solid lines) and predicted responses (open circles) obtained from the provided design parameters. ED (red) and MD (black) extinction cross section spectra are independently expressed. The calculated MSE between predicted and target spectra are (a) 2.64E-3 (b) 1.10E-4 (c) 1.36E-4 and (d) 2.42E-3. Design parameters of the results are summarized in Appendix A2} 
    \label{fig:test-result}
\end{figure}
    
    Total data containing more than 18,000 set are divided into training ({$\sim$}80\%), validating ({$\sim$}10\%), and test data-set ({$\sim$}10\%). NNs are trained using the training data-set, and the trained networks are evaluated by the validation data-set on every epoch to avoid over-fitting problem. After 20,000 epochs, the trained network is finally evaluated by the test data-set that has never been used in previous training or validation steps. Four examples of randomly chosen test results (Fig.~\ref{fig:test-result}) show that the DN is well trained to provide appropriate design parameters of materials and thickness for given input ED and MD responses. The ED and MD responses obtained from predicted designs show good agreement with target input spectra. The average MSE between predicted and target responses of 2,313 test data-set is about 9.43E-3, quantitatively showing that DNs indeed provide appropriate designs that has desired optical properties.
    
    DL-based inverse design is practically utilized to find a structure, which reconstructs an input spectrum with specific purposes. In the following section, we demonstrate the capability of our network to inversely design spherical core-shell nanoparticles, specifically for dipole resonance engineering.


\subsection{Spectral tuning electric dipole resonances}   
\begin{figure}[!ht]
    \centering
    \includegraphics[width=\linewidth]{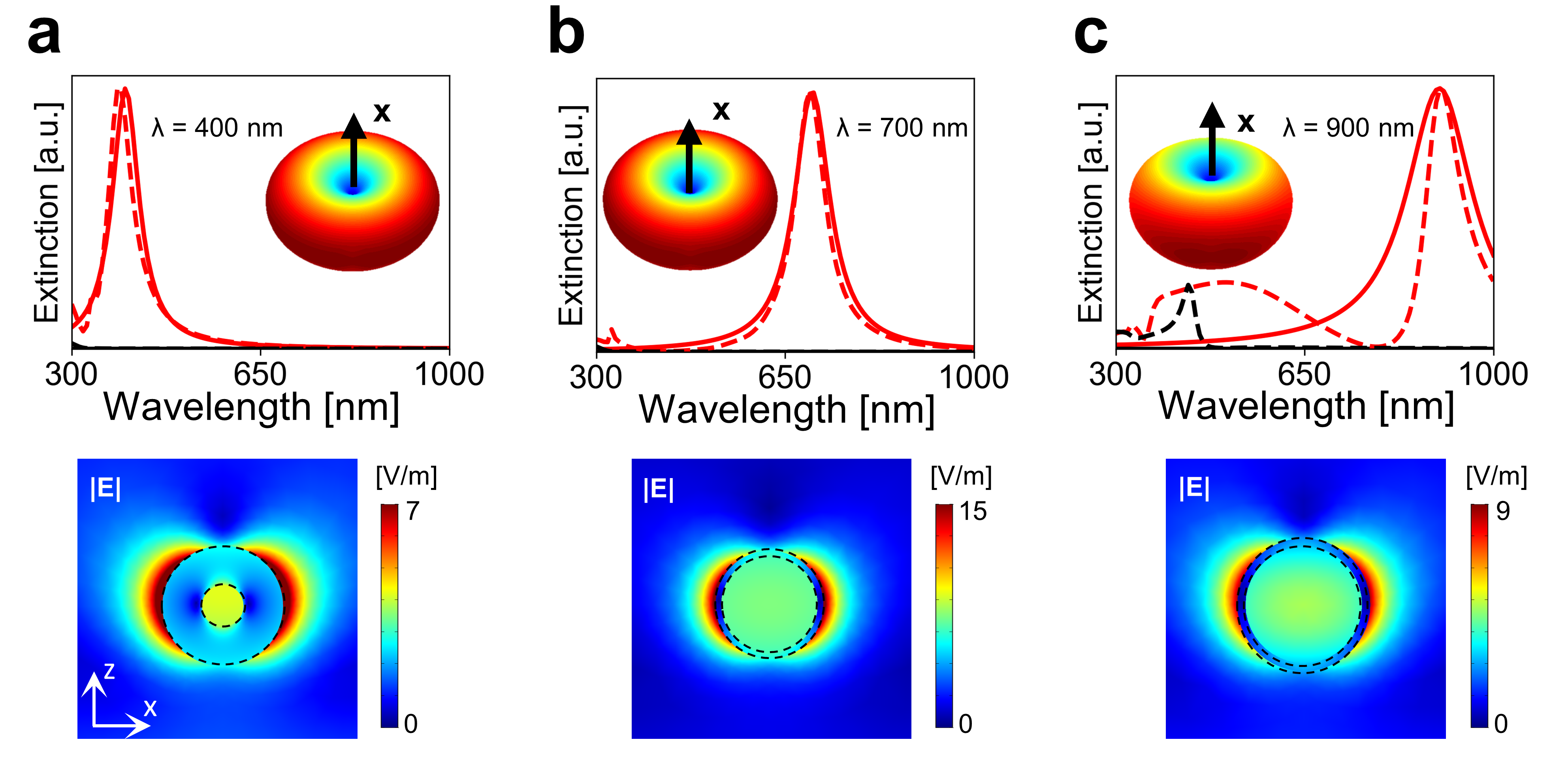}
    \caption{\textbf{Spectrally tuned ED resonances.} Spectra of user-drawn inputs (solid lines) and predicted responses (open circles) obtained from the provided design parameters; ED (red) and MD (black). Insets and bottoms are radiation patterns and electric field distributions at resonant wavelengths of (a) 400~nm, (b) 700~nm, and (c) 900~nm.}
    \label{fig:ED}
\end{figure}

    Sharp extinction resonance peaks of plasmonic nanoparticles have enabled highly sensitive sensing and detection, but resonance wavelength of a single subwavelength nanoparticle is limited by plasmon resonance wavelength of the plasmonic material \cite{willets:2007, mock:2003}. Because choice of plasmonic material and host medium is limited, we believe sensing applications would benefit from spectrally tuned sharp ED resonances, which our DN can achieve for various target wavelengths (Fig.~\ref{fig:ED}). We further confirmed that the found mode is ED mode from field pattern and radiation pattern (Fig.~\ref{fig:ED}).
    
\subsection{Finding spectrally isolated MD resonances} 
\begin{figure}[!ht]
    \centering
    \includegraphics[width=\linewidth]{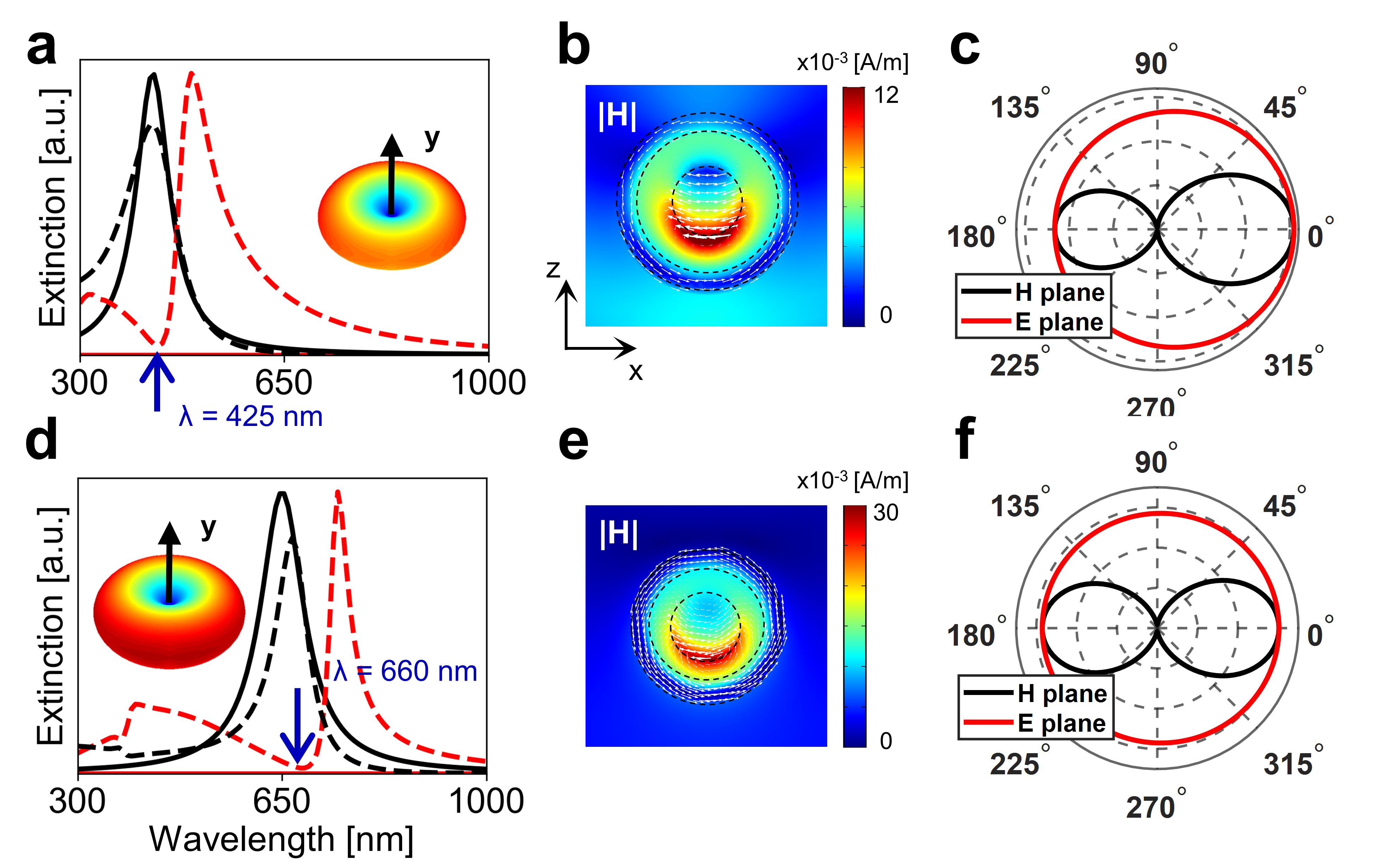}
    \caption{\textbf{Spectrally isolated MD resonance}. (a,d) Spectra of user-drawn inputs (solid lines) and predicted responses (open circles); ED (red) and MD (black). Insets are three dimensional radiation patterns. (b,e) Magnetic field distribution together with the polarization (white arrow) and (c,f) scattering radiation patterns at resonant wavelengths of (b,c) 425~nm and (e,f) 660~nm.}
    \label{fig:MD}
\end{figure}

    In general, subwavelength nanoparticles exhibit dominantly ED mode at planewave incidence. Recently, strong higher-order multipole modes, such as MD and electric quadrupole, have been observed in subwavelength structured plasmonic nanoparticles \cite{zhang:2008}, as well as subwavelength high-index dielectric nanoparticles \cite{evlyukhin:2012}. However, it has been difficult to observe spectrally isolated MD mode from subwavelength nanoparticles \cite{feng:2017}. Only recently, Feng et al. \cite{feng:2017} have founded that a pure MD mode at a certain wavelength where MD mode are overlapped with a non-radiated mode of anapole mode. Such a nanoparticle with spectrally isolated MD may allow us to study light-matter interaction involving MD mode at the absence of ED mode. Such spectrally isolated MD would enable controlled experiments on optomechanical effects of MD \cite{boyer:1988, vaidman:1990}, optical interaction between MDs \cite{yung:1998}, and optical interaction between MD and quantum resonators (i.e.~molecules) and their modified emission properties \cite{schmidt:2012, decker:2013, zambrana:2015, baranov:2017}.
    
    We searched for spectrally isolated MD mode by using input spectra of Lorentzian MD resonance and flat-zero ED spectrum (Fig.~\ref{fig:MD}). We could not find a condition where modes other than MD are completely eliminated over a broad range of wavelength, but we could still find a condition where MD mode is dominant at a certain wavelength ({$\sim$}660~nm in Fig.~\ref{fig:MD}d). This mode is almost purely MD mode as can be seen from its radiation pattern (Fig.~\ref{fig:MD}f). It should be noted that it is not always possible to find structures with physically constrained optical properties, possibly because of limited design conditions (see Appendix D). 
    
\subsection{Finding spectrally overlapped ED and MD resonances} 
\begin{figure}[!ht]
    \centering
    \includegraphics[width=\linewidth]{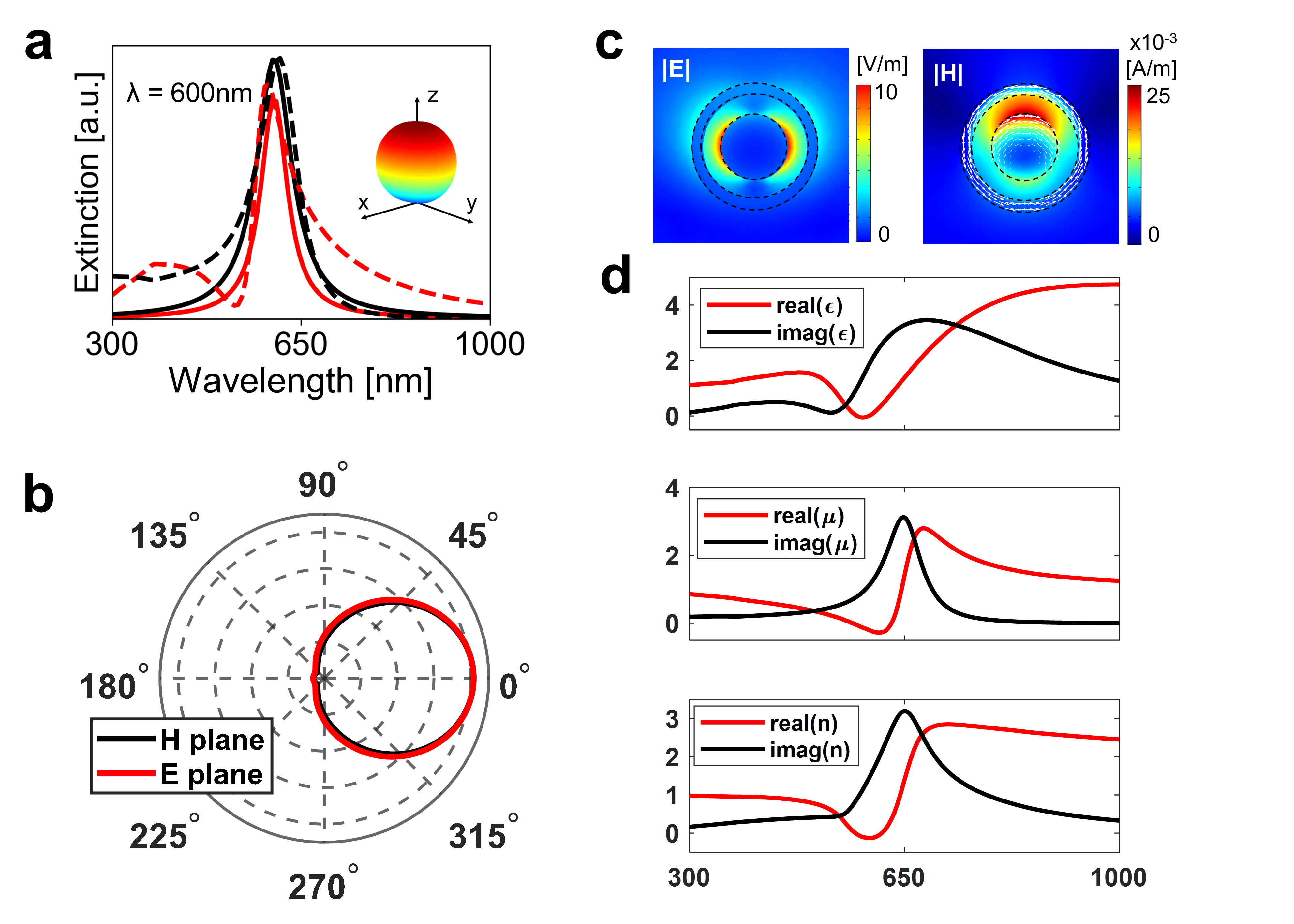}
    \caption{\textbf{Spectrally overlapped ED and MD resonances.} (a) Input spectra (solid lines) and predicted responses (open circles); ED (red) and MD (black). (b) The scattering radiation pattern and (c) electric ($\mathbf{|E|}$) and magnetic ($\mathbf{|H|}$) field distribution at the wavelength of 600~nm. (d) Effective parameters of permittivity (top), permeability (middle), and refractive index (bottom) of the provided core-shell nanoparticle with filling fraction $f = 0.7405$}
    \label{fig:ED-MD}
\end{figure}
    
    Simultaneous ED and MD modes have demonstrated several interesting phenomena including directional scattering \cite{kerker:1983, liu:2012, li:2015, naraghi:2015} and negative index media \cite{paniagua:2011}. However, spectral tuning of both ED and MD resonance to a single target wavelength is difficult, if not cumbersome. Such simultaneous spectrally-tuned ED and MD resonances allow several interesting phenomena including directional scattering and negative index media.
    
    There have already been many publications achieving directional scattering using core-shell nanoparticles \cite{liu:2012, li:2015, naraghi:2015}; however, those publications provide a set of fixed conditions working at certain wavelength. DL-assisted engineering would substantially reduce time and effort taken to find such condition. We could successfully find conditions with simultaneous ED and MD modes using our network (Fig.~\ref{fig:ED-MD}a) with near-zero backward scattering (Fig.~\ref{fig:ED-MD}b). Nanoparticles with large scattering cross-section without backward-scattering would be useful for many photonic devices that require long photon trapping time near acceptor, such as solar cells. Under dipole approximation, zero backward-scattering occurs when ED and MD are in phase (the first Kerker condition), and zero forward-scattering occurs when ED and MD are out of phase (the second Kerker condition) \cite{kerker:1983}. This principle has made high-efficiency metasurface \cite{zheng:2015, decker:2015} and perfect absorber \cite{alaee:2017} possible. 
    
    Another interesting phenomena that would benefit from DL-assisted engineering is design of negative index media. To obtain negative index media, ED and MD resonances need to be simultaneously tuned to a target wavelength. To overcome this difficulty, Pendry proposed chiral media as a possible route to negative index media without simultaneously engineering ED and MD resonances \cite{pendry:2004}. Our designed core-shell nanoparticle exhibits negative refractive index $n=-0.118+1.684i$, as well as simultaneously negative effective permittivity and permeability, at the target wavelength of 600~nm (Fig.~\ref{fig:ED-MD}c). 
    Here, effective permittivity and permeability are obtained using Clausius-Mossotti relation (Eqn.~\ref{eqn:C-M}):
    \begin{equation}
        \frac{\epsilon_{\text{eff}}-\epsilon_0}{\epsilon_{\text{eff}}+2\epsilon_0} = f\frac{\alpha_E}{4\pi R^3}, \frac{\mu_{\text{eff}}-\mu_0}{\mu_{\text{eff}}+2\mu_0} = f\frac{\alpha_M}{4\pi R^3},
        \label{eqn:C-M}
    \end{equation}
    where $\alpha_E$, and $\alpha_M$ are the electric and magnetic polarizabilities.
    
\subsection{Discussion} 
    Multiplexed input spectra with different polarizations have been demonstrated in DL \cite{malkiel2018plasmonic, liu2018generative}; however, multiplexed input spectra with multipole-decomposed spectra have not been reported. We believe this approach could be used to directly engineer many physically intriguing phenomena, because simple extinction spectra cannot give deeper physical insights, which the multipole-decomposed spectra may provide. In this present study, we limited the input spectra to only ED and MD mode. By additionally introducing higher-order modes to the input spectra, many designs with interesting phenomena may be designed as well. However, it is not always possible to find a structure with an arbitrary spectrum, possibly because of limited geometry of the three-layered spherical core-shell nanoparticles and the size limit. If target application is fixed, it should be better to directly use optical properties, such as directionality for directional scattering and effective refractive index for negative index media, instead of general extinction spectra. Nevertheless, this work shows great potential of DL-based inverse design by introducing materials as parameters and thereby significantly extending the degree-of-freedom of design possibility. Also, as data throughput increases, it is expected that many new designs will be available that were difficult to design empirically or intuitively. 
    
\section{Conclusion}
    In conclusion, we have provided the first simultaneous inverse design of material and structural parameters of core-shell nanoparticles based on independently given ED and MD spectra. Materials are indexed by numbering and designed by DL, along with the structural parameters of the thicknesses. This combined classification (material) and regression (thickness) problem is solved by our devised loss function. To show the capability of our DL network, we specifically demonstrate DL-assisted inverse design for three dipole resonance engineering problems with on-demand applications using spherical core-shell nanoparticles: 1) spectral tuning of ED resonances, for sensing applications; 2) finding spectrally isolated MD resonances; and 3) finding spectrally overlapped ED and MD resonances, for directional scattering or negative index media. We present individual engineering of ED and MD spectra to discuss their relevance to physical phenomena. We believe that our concept of DL-assisted simultaneous inverse design of materials and structures will lead to the development of the field of nanophotonics by allowing efficient inverse design of nanophotonic structures. 

\begin{acknowledgments}
This work is financially supported by the national Research Foundation grants (NRF-2019R1A2C3003129, CAMM-2019M3A6B3030637, NRF-2018M3D1A1058998, \& NRF-2015R1A5A1037668) funded by the Ministry of Science and ICT, Korea. S.S acknowledges global Ph.D fellowship (NRF-2017H1A2A1043322) from the NRF-MSIT, Korea.
\end{acknowledgments}

\appendix
\setcounter{figure}{0}
\setcounter{table}{0}
\renewcommand{\thetable}{A\arabic{table}}
\renewcommand{\thefigure}{A\arabic{figure}}
\section{Details on deep-learning procedure}
\begin{figure}[h]
        \includegraphics[width=\linewidth]{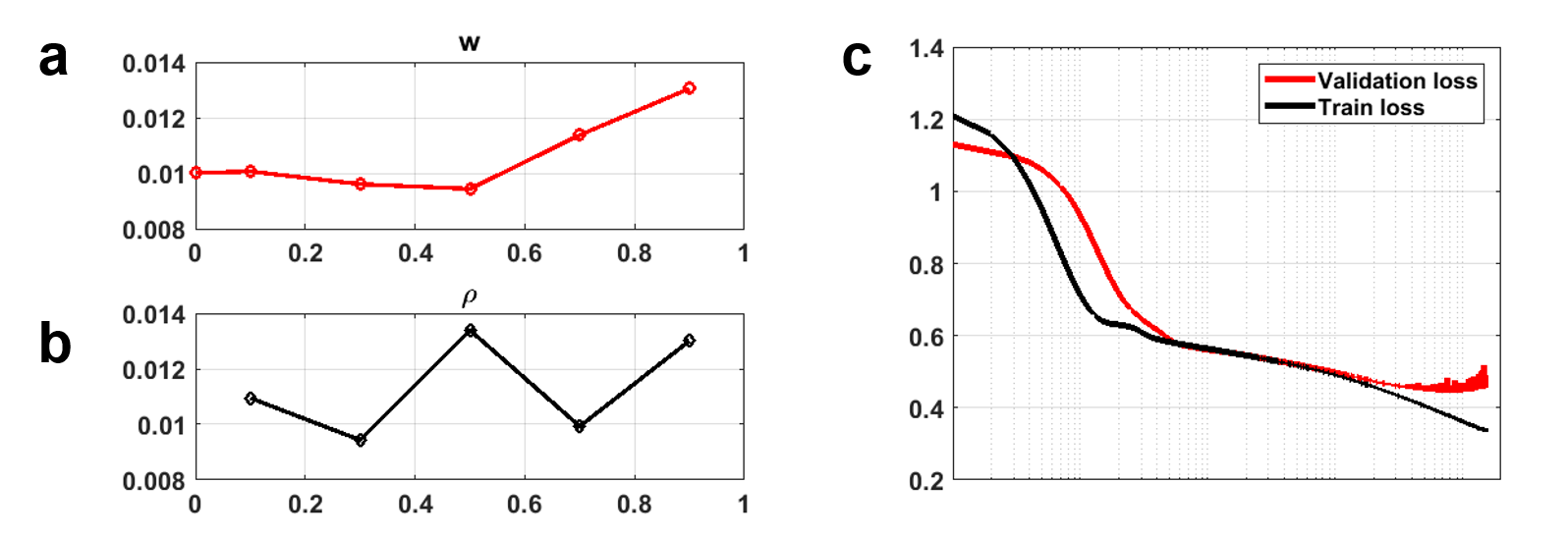}
        \centering
        \caption{MSE errors as a function of (a) $w$ and (b) $\rho$. Values of $w = 0.3$ and $\rho = 0.5$ minimize the average MSE error between input and predicted spectra of the test dataset, and therefore chosen for training. (c) Learning curves, which show the MSE errors as a function of training iterations, of the training (black) and validation (red) dataset. Logarithmic scale is used on the horizontal-axis.}
        \label{fig:deep learning result}
    \end{figure}

    We have developed a weighted loss function to solve simultaneous inverse design of materials and structure parameters. Therefore, we conduct parametric study for two different weight of w and $\rho$ as discussed in the main text (Fig.~\ref{fig:deep learning result}a,b). 
    
    After each 20,000 epochs of training using different weights, networks are converged to different states depending on different values of w and $\rho$. The trained networks are quantitatively evaluated using test dataset by calculating MSE between input spectra and predicted spectra obtained from the DL-assisted designs. Values of $w = 0.5$ and $\rho = 0.3$ minimize the average MSE error between input and predicted spectra of the test dataset, and therefore chosen for training. The learning curve of the optimized network shows that DN is well trained with the converged losses after 20,000 epochs for each validation and train data (Fig.~\ref{fig:deep learning result}c). Table~\ref{tab:hyperparameters} summarizes other hyper-parameters for training. We use Pytorch framework for DL, and batch gradient descent method is used. The network is trained on an instance with single NVIDIA GTX-1080Ti, and it takes approximately 100 minutes to carry out 20,000 epochs. Loss criteria discussed in the main text are given as:
    \begin{align}
        l_{\text{MSE}}(x,y) &= \frac{1}{n} \sum_{n} (x_n-y_n)^2 \\
        l_{\text{MAE}}(x,y) &= \frac{1}{n} \sum_{n} |x_n-y_n| \\ 
        l_{\text{BCE}}(x,y) &= -[y\log \sigma (x)+(1-y)\log(1-\sigma(x))]
        \label{eqn:loss crieria}
    \end{align}
    Here, $x$ and $y$ are the target and the output, respectively, and $\sigma(x)$ is Sigmoid function. 
    \begin{table}[h]
        \small
        \resizebox{\linewidth}{!}{%
        \begin{tabular}{c|cc}
        \hline
        & DN & SN \\
        \hline
        Optimizer & \makecell{SGD \\ (lr = 0.01, momentum = 0.9)} & \makecell{Adam \\ (lr = 5E-4, weight decay = 1E-5)}\\
        Activation Function & ReLU & leakyReLU, $f(x)=\max{(0.01x,x)}$\\
        The number of neurons & [500, 2000, 2000, 100] & [100, 1500, 1500, 300] \\
        \hline
        \end{tabular}}
        \centering
        \caption{\textbf{Hyperparameters for deep learning} lr stands for learning rate.}
        \label{tab:hyperparameters}
    \end{table}
    
\section{Design parameters used in the main text}
    Design parameters of the results discussed in the main text are summarized in Table.~\ref{tab:Design parameters in the main text}. As can be seen in Table.~\ref{tab:Design parameters in the main text}, negative values of thickness are returned, which are physically nonsense. However, it should be noted that such negative values occur only if the corresponding materials are 0 (none). In previous training steps, we used 0 (none) material to represent one or two-layered core-shell nanoparticles. DNs may have learned through the training that the thickness values correseponding to 0 (none) material does not affect to the optical properties. To avoid negative values of thickness, activation functions of the last neurons can be replaced to other activation functions of ReLU or Sigmoid that provides only positive outputs. 
    
    \begin{table}[h]
        \small
        \centering
        \resizebox{\linewidth}{!}{%
        \begin{tabular}{c|cccccc}
        \hline
        Case & $m_1$ & $m_2$ & $m_3$ & $t_1$ [nm] & $t_2$ [nm] & $t_3$ [nm] \\
        \hline
        Fig.2a & 4 (Cu) & 4 (Cu) & 7 (Si) & 10.45 & 19.72 & 58.27 \\
        Fig.2b & 1 (Ag) & 2 (Au) & 7 (Si) & 33.07 & 96.74 & 19.57 \\
        Fig.2c & 3 (Al) & 4 (Cu) & 0 (none) & 119.39 & 20.01 & -0.46 \\
        Fig.2d & 4 (Cu) & 7 (Si) & 5 (TiO$_2$) & 28.97 & 20.00 & 80.56 \\
        ED @ $\lambda = 400 \text{nm}$ (Fig.3a) & 6 (SiO$_2$) & 1 (Ag) & 0 (none) & 15.49 & 29.41 & 1.26 \\
        ED @ $\lambda = 700 \text{nm}$ (Fig.3b) & 6 (SiO$_2$) & 1 (Ag) & 0 (none) & 57.57 & 6.72 & -2.85 \\
        ED @ $\lambda = 900 \text{nm}$ (Fig.3c) & 5 (TiO$_2$) & 1 (Ag) & 0 (none) & 91.12 & 12.81 & 11.02 \\
        MD @ $\lambda = 425 \text{nm}$ (Fig.4a) & 1 (Ag) & 6 (SiO$_2$) & 5 (TiO$_2$) & 37.97 & 39.25 & 19.65 \\
        MD @ $\lambda = 650 \text{nm}$ (Fig.4b) & 1 (Ag) & 5 (TiO$_2$) & 7 (Si) & 42.52 & 28.79 & 25.67 \\
        ED \& MD @ $\lambda = 600 \text{nm}$ (Fig.5a) & 1 (Ag) & 6 (SiO$_2$) & 7 (Si) & 53.98 & 30.38 & 18.41 \\
        \hline
        \end{tabular}}
        \caption{\textbf{Design parameters of predicted structures of the Fig.{2$\sim$5} in the main text.}}
        \label{tab:Design parameters in the main text}
    \end{table}
\section{Additional examples not included in the main text}
    \begin{figure}[!ht]
        \centering
        \includegraphics[width=\linewidth]{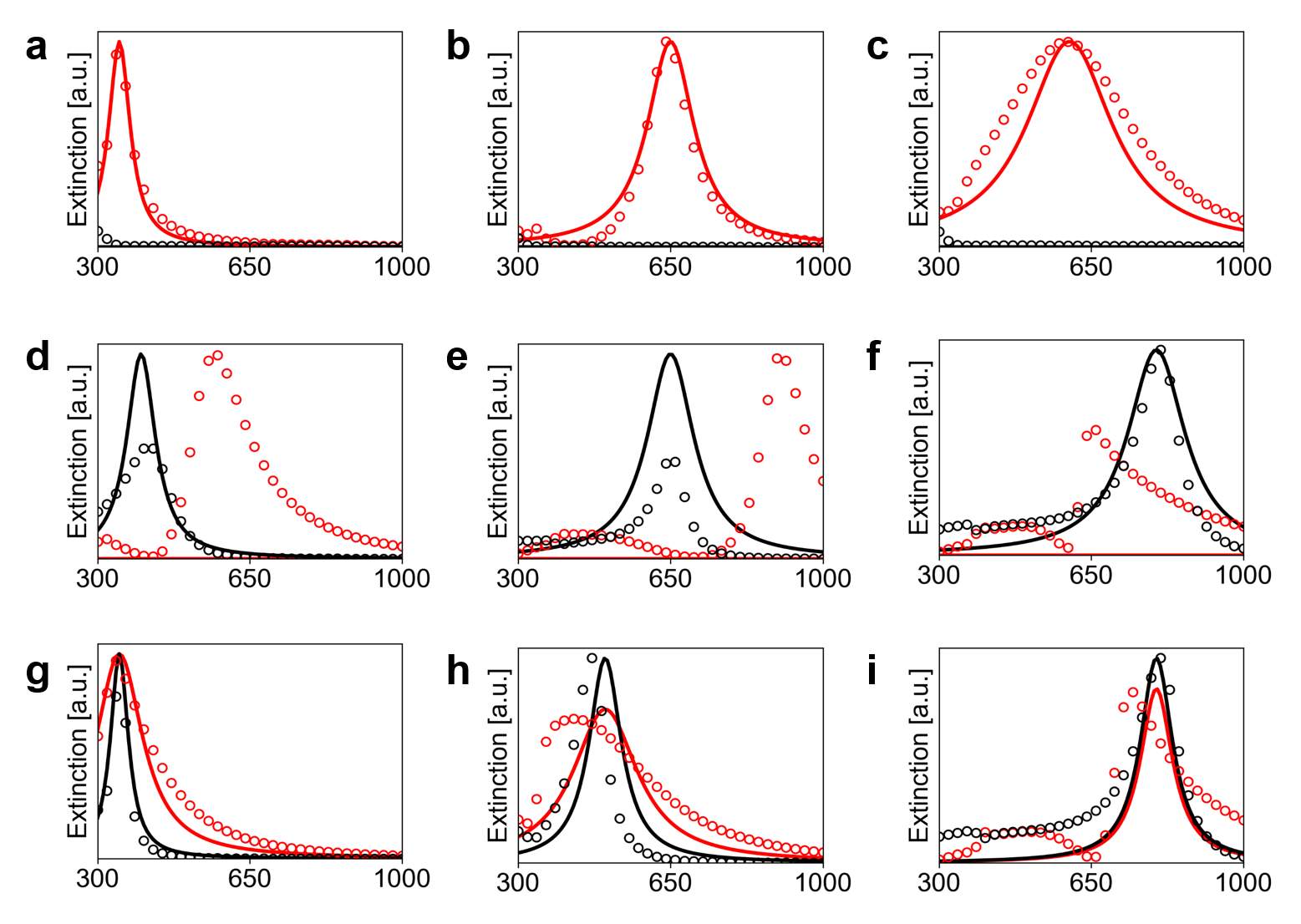}
        \caption{\textbf{Additional examples for on-demand application} Additional examples are shown for (a)-(c) three ED resonant modes, (d)-(f) three MD resonant modes, and (g)-(i) spectrally overlapped ED and MD modes at various wavelengths. Design parameters of each cases are summarized in Table~\ref{tab:Design parameters in SI}.} 
        \label{fig:additional-result}
    \end{figure}
    \begin{table}[h]
        \small
        \centering
        \resizebox{\linewidth}{!}{%
        \begin{tabular}{c|cccccc}
        \hline
        Case & $m_1$ & $m_2$ & $m_3$ & $t_1$ [nm] & $t_2$ [nm] & $t_3$ [nm] \\
        \hline
        ED @ $\lambda = 350 \text{nm}$ (Fig.S1a) & 3 (Al) & 5 (TiO$_2$) & 0 (none) & 33.33 & 10.82 & 1.88 \\
        ED @ $\lambda = 650 \text{nm}$ (Fig.S1b) & 6 (SiO$_2$) & 1 (Ag) & 0 (none) & 63.03 & 9.98 & 2.01 \\
        ED @ $\lambda = 600 \text{nm}$ (Fig.S1c) & 3 (Al) & 1 (Ag) & 6 (SiO$_2$) & 60.73 & 21.55 & 28.01 \\
        MD @ $\lambda = 400 \text{nm}$ (Fig.S1d) & 1 (Ag) & 6 (SiO$_2$) & 5 (TiO$_2$) & 57.19 & 35.04 & 15.78 \\
        MD @ $\lambda = 650 \text{nm}$ (Fig.S1e) & 1 (Ag) & 1 (Ag) & 7 (Si) & 46.20 & 36.50 & 31.71 \\
        MD @ $\lambda = 800 \text{nm}$ (Fig.S1f) & 1 (Ag) & 6 (SiO$_2$) & 7 (Si) & 58.41 & 43.87 & 30.25 \\
        ED \& MD @ $\lambda = 350 \text{nm}$ (Fig.S1g) & 5 (TiO$_2$) & 6 (SiO$_2$) & 0 (none) & 50.60 & 35.46 & -6.91 \\
        ED \& MD @ $\lambda = 500 \text{nm}$ (Fig.S1h) & 6 (SiO$_2$) & 5 (TiO$_2$) & 6 (SiO$_2$) & 32.92 & 50.35 & 29.14 \\
        ED \& MD @ $\lambda = 800 \text{nm}$ (Fig.S1i) & 1 (Ag) & 6 (SiO$_2$) & 7 (Si) & 71.86 & 29.15 & 33.23 \\
        \hline
        \end{tabular}}
        \centering
        \caption{\textbf{Design parameters of predicted structures in Fig.~\ref{fig:additional-result}}}
        \label{tab:Design parameters in SI}
    \end{table}
\section{Inverse design using arbitrary input spectra}
    
    The spectral forms of electromagnetic resonances can generally be regarded as Lorentzian shape. Therefore, we employed Lorentzian-type spectra to describe dipole resonances throughout the whole manuscript as shown in Eqn.~\ref{eqn:Lorentzian}. Parameters used in the main text to generate Lorentzian-shape spectra are summarized in Table.~\ref{tab:Lorentzian parameters}.  
    
    \begin{equation}
        f(x;a,b,c) = \frac{ac^2}{(x-b)^2+c^2}
        \label{eqn:Lorentzian}
    \end{equation}
    
    \begin{table}[h]
        \small
        \centering
        \resizebox{\linewidth}{!}{%
        \begin{tabular}{c|cccccc}
        \hline
        Case & a$_{\text{ed}}$ & b$_{\text{ed}}$ & c$_{\text{ed}}$ & a$_{\text{md}}$ & b$_{\text{md}}$ & c$_{\text{md}}$ \\
        \hline
        ED @ $\lambda = 400 \text{nm}$ (Fig.3a) & 1 & 400 & 30 & 0 & 0 & 0 \\
        ED @ $\lambda = 700 \text{nm}$ (Fig.3b) & 1 & 700 & 40 & 0 & 0 & 0 \\
        ED @ $\lambda = 900 \text{nm}$ (Fig.3c) & 1 & 900 & 70 & 0 & 0 & 0 \\
        MD @ $\lambda = 425 \text{nm}$ (Fig.4a) & 0 & 0 & 0 & 1 & 425 & 35 \\
        MD @ $\lambda = 650 \text{nm}$ (Fig.4b) & 0 & 0 & 0 & 1 & 650 & 45 \\
        ED \& MD @ $\lambda = 600 \text{nm}$ (Fig.5a) & 0.85 & 600 & 30 & 1 & 600 & 40 \\
        \hline
        \end{tabular}}
        \centering
        \caption{\textbf{Lorentzian parameters used in the main text}}
        \label{tab:Lorentzian parameters}
    \end{table}
    
\begin{figure}[!ht]
        \centering
        \includegraphics[width=\linewidth]{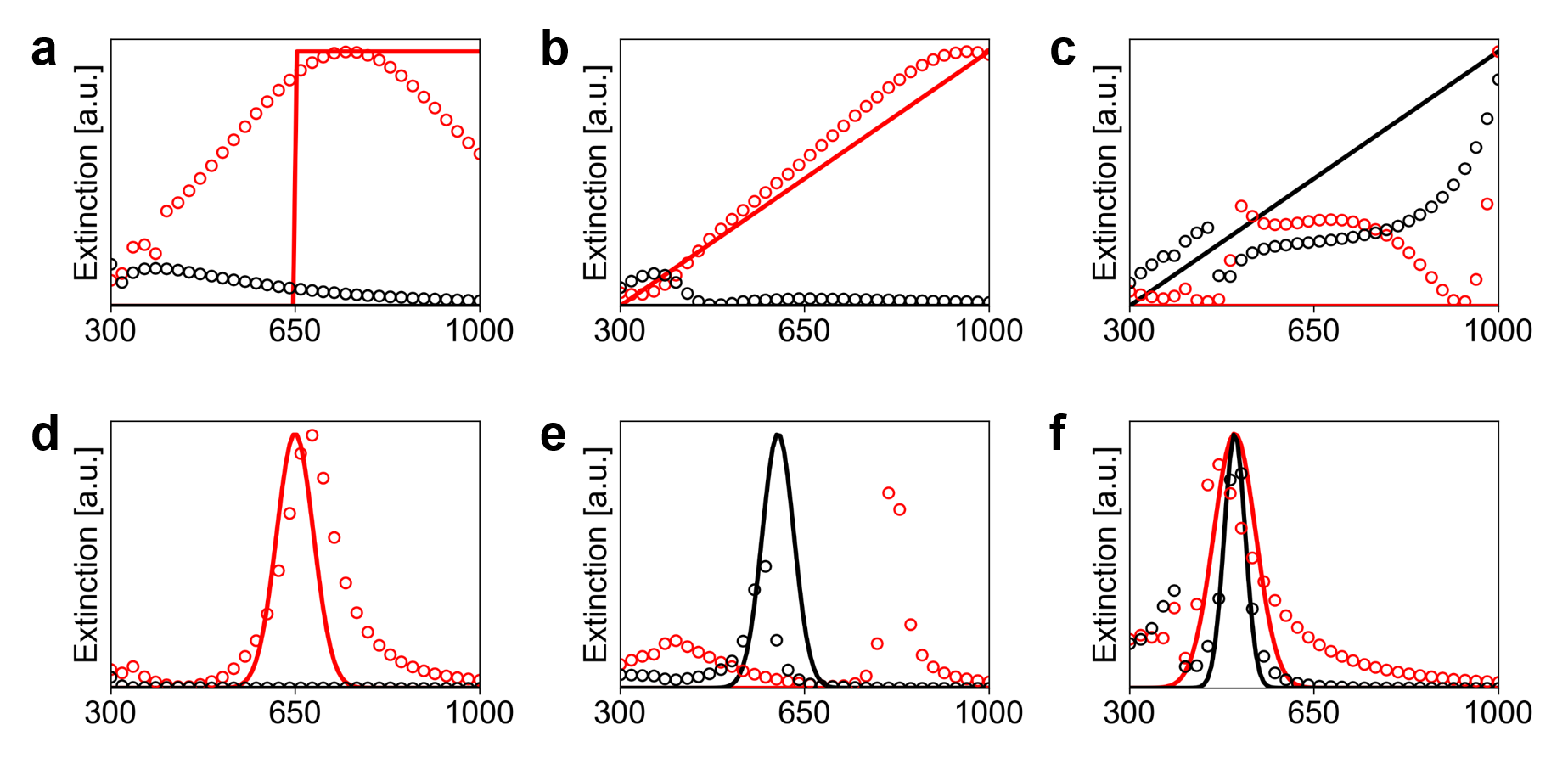}
        \caption{\textbf{Additional examples for random spectra other than Lorentzian resonances} Design parameters of each cases are summarized in Table~\ref{tab:Design parameters in SI2}.} 
        \label{fig:random-spectra}
    \end{figure}

    We further test our network with random spectra with step function, linear function, and Gaussian resonances \ref{fig:random-spectra}. For a user-drawn spectrum with non-physical shape, the network is likely to fail to generate appropriate design that have desired input spectra. 
    
    \begin{table}[h]
        \small
        \centering
        \resizebox{\linewidth}{!}{%
        \begin{tabular}{c|cccccc}
        \hline
        Case & $m_1$ & $m_2$ & $m_3$ & $t_1$ [nm] & $t_2$ [nm] & $t_3$ [nm] \\
        \hline
        ED step function (Fig.S3a) & 6 (SiO$_2$) & 1 (Ag) & 0 (none) & 69.82 & 62.53 & 20.56 \\
        ED linear function (Fig.S3b) & 3 (Al) & 1 (Ag) & 5 (TiO$_2$) & 100.33 & 45.77 & 23.74 \\
        MD linear function (Fig.S3c) & 1 (Ag) & 5 (TiO$_2$) & 7 (Si) & 64.94 & 63.83 & 43.17 \\
        ED Gaussian function (Fig.S3d) & 6 (SiO$_2$) & 1 (Ag) & 0 (No) & 66.6 & 9.15 & -3.10 \\
        MD Gaussian function (Fig.S3e) & 1 (Ag) & 1 (Ag) & 7 (Si) & 46.54 & 35.69 & 27.48 \\
        ED MD Gaussian function (Fig.S3f) & 7 (Si) & 5 (TiO$_2$) & 0 (none) & 44.44 & 41.13 & 18.49 \\
        \hline
        \end{tabular}}
        \centering
        \caption{\textbf{Design parameters of predicted structures in Fig.~\ref{fig:random-spectra}}}
        \label{tab:Design parameters in SI2}
    \end{table}
\section{Response to gradual change of input spectra}

\begin{figure}[h]
    \includegraphics[width=\linewidth]{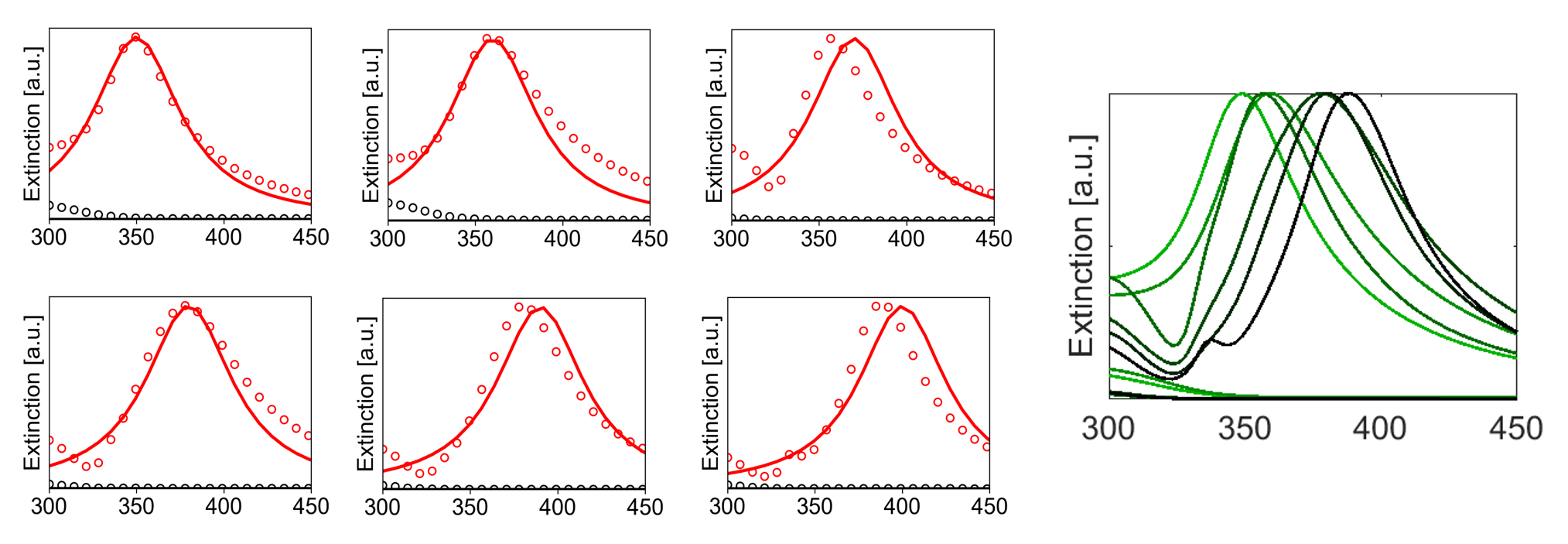}
    \centering
    \caption{Response to gradual change of input spectra}
    \label{fig:response to gradual change of the input spectra}
    \end{figure}

    We explore the transition of the designed material using gradually tuned input spectra (Fig.~\ref{fig:response to gradual change of the input spectra}). By changing resonant wavelengths of ED extinction spectra from $\lambda=350$~nm to $\lambda=400$~nm, suggested designs by DL are also changed (Table.~\ref{tab:Design parameters in SI3}). At the shorter resonant wavelength, materials of Al with higher plasma frequency ($\lambda_\text{plasma}\sim 80$nm) is returned. By increasing the resonant wavelength, material transition between Al to Ag with relatively lower plasma frequency ($\lambda_\text{plasma}\sim 130$nm) are shown. This suggests that DN has successfully learned the mapping between material information of plasmonic resonances and optical properties well, even if we used material information with the simplified indexed numbers. 

    \begin{table}[h]
        \small
        \centering
        \resizebox{\linewidth}{!}{%
        \begin{tabular}{c|cccccc}
        \hline
        Case & $m_1$ & $m_2$ & $m_3$ & $t_1$ [nm] & $t_2$ [nm] & $t_3$ [nm] \\
        \hline
        ED @ $\lambda = 350 \text{nm}$ & 3 (Al) & 5 (TiO$_2$) & 0 (none) & 32.72 & 9.24 & -1.16 \\
        ED @ $\lambda = 360 \text{nm}$ & 3 (Al) & 5 (TiO$_2$) & 0 (none) & 36.32 & 10.16 & -0.88 \\
        ED @ $\lambda = 370 \text{nm}$ & 3 (Al) & 1 (Ag) & 0 (none) & 30.95 & 11.95 & 1.49 \\
        ED @ $\lambda = 380 \text{nm}$ & 2 (Au) & 1 (Ag) & 0 (none) & 22.18 & 19.25 & 3.17 \\
        ED @ $\lambda = 390 \text{nm}$ & 3 (Al) & 1 (Ag) & 0 (none) & 18.67 & 26.12 & 1.40 \\
        ED @ $\lambda = 400 \text{nm}$ & 6 (SiO$_2$) & 1 (Ag) & 0 (none) & 13.49 & 30.54 & 0.50 \\
        \hline
        \end{tabular}}
        \caption{\textbf{Design parameters of predicted structures in Fig.~\ref{fig:response to gradual change of the input spectra}}}
        \label{tab:Design parameters in SI3}
    \end{table}
\section{Extinction from spherical core-shell nanoparticles}
    Extinctions from spherical core-shell nanoparticles were efficiently calculated using the T-matrix method, which is an exact semi-analytical method based on vector spherical waves. The incident and scattered field for a nanoparticle is expanded using vector spherical wave functions as basis:
\begin{align}
    \mathbf{E}_{inc} &= E_0\sum_{nm}{
        a_{nm}^M\mathbf{M}_{nm}^{1}(k\mathbf{r})
        +a_{nm}^E\mathbf{N}_{nm}^{1}(k\mathbf{r})},\\
    \mathbf{H}_{inc} &= \frac{E_0}{i\eta}\sum_{nm}{
        a_{nm}^M\mathbf{N}_{nm}^{1}(k\mathbf{r})
        +a_{nm}^E\mathbf{M}_{nm}^{1}(k\mathbf{r})},\\
    \mathbf{E}_{sca} &= E_0\sum_{nm}{
        b_{nm}^M\mathbf{M}_{nm}^{3}(k\mathbf{r})
        +b_{nm}^E\mathbf{N}_{nm}^{3}(k\mathbf{r})}, \\
    \mathbf{H}_{sca} &= \frac{E_0}{i\eta}\sum_{nm}{
        b_{nm}^M\mathbf{M}_{nm}^{3}(k\mathbf{r})
        +b_{nm}^E\mathbf{N}_{nm}^{3}(k\mathbf{r})},
\end{align}
    where $a_{nm}^{E,M}$ and $b_{nm}^{E,M}$ are incident and scattered multipoles with superscripts $E$ and $M$ representing electric and magnetic modes; superscripts $1$ and $3$ represents regular and singular vector spherical wave functions, respectively; $k$ is the wavenumber and $\eta$ is the wave impedance in host medium. The summation index $n$ goes from 1 to infinity, but needs to be truncated by $n_\mathrm{max}$ until convergence is reached, and index $m$ goes from $-n$ to $n$. In this work, we use $n_\mathrm{max}=1$ to consider only dipole.
    
    The incident and scattered multipoles are linearly related by the T-matrix:
\begin{equation}
    \begin{pmatrix}b_{nm}^E\\b_{nm}^M\end{pmatrix}
    = T
    \begin{pmatrix}a_{nm}^E\\a_{nm}^M\end{pmatrix}
    = \begin{pmatrix} T_{ee} & T_{em} \\ T_{me} & T_{mm} \end{pmatrix}
    \begin{pmatrix}a_{nm}^E\\a_{nm}^M\end{pmatrix}.
\end{equation}
    T-matrix of spherical core-shell nanoparticles can be calculated by setting the fields inside \textit{j}-th shell as:

\begin{eqnarray}
    \mathbf{E}_{j} = E_0\sum_{nm}[
        A_{j,nm}^M\mathbf{M}_{nm}^{1}(k_j\mathbf{r})
        +A_{j,nm}^E\mathbf{N}_{nm}^{1}(k_j\mathbf{r})\nonumber \\
        +B_{j,nm}^M\mathbf{M}_{nm}^{3}(k_j\mathbf{r})
        +B_{j,nm}^E\mathbf{N}_{nm}^{3}(k_j\mathbf{r})]\,
\end{eqnarray}
\begin{eqnarray}
    \mathbf{H}_{j} = \frac{E_0}{i\eta_j}\sum_{nm}[
        A_{j,nm}^M\mathbf{N}_{nm}^{1}(k_j\mathbf{r})
        +A_{j,nm}^E\mathbf{M}_{nm}^{1}(k_j\mathbf{r})\nonumber \\
        +B_{j,nm}^M\mathbf{N}_{nm}^{3}(k_j\mathbf{r})
        +B_{j,nm}^E\mathbf{M}_{nm}^{3}(k_j\mathbf{r})]\,
\end{eqnarray}
    where $k_j$ is the wavenumber and $\eta_j$ is the wave impedance inside the $j$-th shell. The unknown coefficients are related by the boundary conditions $\mathbf{\hat{r}}\times\mathbf{E}_{in} = \mathbf{\hat{r}}\times\mathbf{E}_{out}$ and $\mathbf{\hat{r}}\times\mathbf{H}_{in} = \mathbf{\hat{r}}\times\mathbf{H}_{out}$ at spherical surface and interfaces, and solved using transfer-matrix method. Note that $B_{j,nm}^{E,M}$ at the core are zero.
    
    Finally, the extinction from from a scatterer is calculated as
\begin{equation}
    P_{ext} = -\frac{E_0^2}{2\eta k^2} \sum_{nm}{\Re{(b_{nm}^E a_{nm}^{E*} +b_{nm}^M a_{nm}^{M*})}},
\end{equation}
    using the incident multipoles for planewave calculated as Ref.~\cite{moreira:2016} and the scattered multipoles calculated using the incident multipoles and T-matrix. This approach is much more computationally efficient, as well as accurate, compared to numerical methods, such as FEM and FDTD, so this T-matrix method is beneficial for dataset generation in DL, if the T-matrix can be easily calculated. Optical properties of Ag and Au were taken from \cite{johnson1972optical}, Al and Cu from \cite{rakic1998optical}, Si from \cite{palik1998handbook}, TiO$_2$ from \cite{siefke2016materials}, and SiO$_2$ from \cite{rodriguez2016self}.

\nocite{*}

\bibliography{apssamp}

\end{document}